\documentclass{article}

\usepackage{arxiv}

\usepackage[utf8]{inputenc} 
\usepackage[T1]{fontenc}    
\usepackage{hyperref}       
\usepackage{url}            
\usepackage{booktabs}       
\usepackage{amsmath,amssymb,amsfonts} 
\usepackage{nicefrac}       
\usepackage{microtype}      
\usepackage{lipsum}
\usepackage{graphicx}
\usepackage{bm}
\graphicspath{ {./images/} }

\newcommand{\dd}{\mathrm{d}}

\usepackage{tikz}

\newcommand\submittedtext{%
  \footnotesize This work has been submitted to the IEEE for possible publication. Copyright may be transferred without notice, after which this version may no longer be accessible.}

\newcommand\submittednotice{%
\begin{tikzpicture}[remember picture,overlay]
\node[anchor=south,yshift=10pt] at (current page.south) {\fbox{\parbox{\dimexpr0.65\textwidth-\fboxsep-\fboxrule\relax}{\submittedtext}}};
\end{tikzpicture}%
}

\title{Method for Mitigating Attention to Inappropriate Content Based on Attention Dynamics Model}

\author{
 Naoki Hirakura \\
  Depertment of Information Systems Engineering\\
  Toyama Prefectural University\\
  5180 Kurokawa, Imizu, Japan\\
  \texttt{hirakura@pu-toyama.ac.jp}
}

\begin{document}
\maketitle
\begin{abstract}
The expansion of the attention economy has led to the growing issue of inappropriate content being posted by profit-driven users.
Previous countermeasures against inappropriate content have relied on moderation, which raises ethical concerns, or information diffusion control, which requires considering larger scale networks, including general users.
This study proposes an imitation strategy as an intervention method that does not rely on moderation and focuses on a relatively smaller scale competitive network of information disseminators rather than the entire social network.
The imitation strategy is a novel approach that utilizes increased competition among information disseminators through imitation to reduce attention to inappropriate content.
Through theoretical analysis and numerical simulations, I demonstrate that the imitation strategy is more effective when nodes with higher eigenvector centrality are selected as targets and nodes with lower eigenvector centrality are chosen as imitators.
\end{abstract}


\section{Introduction}
\submittednotice
The attention economy has expanded alongside the widespread adoption of social media\cite{attention_economy_1997,attention_economy_2001,attention_economy_2021}. 
The attention economy is a context in which attention holds economic value. 
A significant portion of the revenue for social media platforms is primarily sourced from advertising, making it crucial to maximize user engagement time. 
To achieve this, social media platforms incentivize popular users to encourage content creation, aiming to maximize user engagement time~\cite{profit_model}.
While this approach leads to increased content creation and provides users with access to a greater volume of content, its negative aspects have become more apparent in recent years. 
Specifically, this has led to the emergence of users who post inappropriate content driven by the mindset that any content is acceptable as long as it generates their profit.
For example, there has been an increase in posts that prioritize impact over truth, such as fake news~\cite{fake_news,fake_news_truefalse}, as well as content deliberately designed to incite online flaming~\cite{online_flaming}. 
The widespread dissemination of inappropriate content on social media has significant negative impacts on society. 
The spread of fake news can influence election results~\cite{fake_news_election}, and the dissemination of false medical information can lead to refusal of vaccines, which can undermine the effectiveness of public health measures~\cite{fake_news_infodemic}.
Acts of attracting attention through online flaming may include hate speech, which can exacerbate real-world conflicts~\cite{hate_speech}.
The recent advancements in generative AI technology have accelerated the creation of inappropriate content, and this issue is expected to become even more severe in the near future~\cite{generativeAI_fake_news}.

Given this background, understanding attention dynamics in the attention economy and suppressing attention to inappropriate content are essential.
Thus far, multiple studies have been conducted on attention dynamics.
For example, the relationship between information novelty and fluctuations in attention \cite{novelty_attention}, patterns of attention decay \cite{universal_decay}, transience of attention \cite{transience}, and attention burst phenomena \cite{burst} have been elucidated.
Research on attention inequality has revealed that on YouTube, 3\% of channels account for 85\% of total views \cite{youtube_attention}, while on Twitter (now X), 20\% of accounts generate more than 90\% of engagement \cite{twitter_attention}.
Several models of attention fluctuation based on the competitive dynamics among content have been proposed.
The competition among memes \cite{meme_competition_popularity,meme_competition} and the competition among topics \cite{attention_model} have been modeled.
Various studies have been conducted from different perspectives, primarily focusing on attention to online content.
However, research on mitigating attention to inappropriate content has been insufficiently explored.

Research aimed at mitigating the impact of inappropriate content on social media is conducted primarily within the framework of the influence minimization problem.
These studies use the independent cascade model and the linear threshold model~\cite{icm}, an information diffusion model, to identify the set of nodes or edges that can most effectively suppress the influence of inappropriate content. 
The purpose of these studies is to determine which nodes or edges should be subject to moderation, such as deletion or restriction of accounts.
For example, research has been conducted on suppressing the spread of negative information by deleting nodes~\cite{minimizing_node_topic}, as well as on suppressing the spread of information by removing edges~\cite{minimizing_edge_1,minimizing_edge_2}.
Additionally, a method has been proposed that minimizes misinformation spread by probabilistically deleting edges between users rather than removing specific edges~\cite{edge_dropout}.
However, several issues have been raised regarding moderation.
In particular, concerns have been raised about the lack of fairness and transparency, as well as the potential for certain opinions or communities to be excessively suppressed~\cite{moderation_issue}.
Excessive reliance on moderation carries the risk of hindering free expression and diversity, making it a limited solution as a countermeasure.

In addition to studies conducted within the framework of influence minimization problems that assume moderation by removing specific nodes or edges, research has also been conducted on influence limitation, which aims to suppress the impact of inappropriate content by effectively disseminating counter information \cite{limiting2011}.
Additionally, methods for containing the spread of misinformation have also been proposed \cite{containment2012, containment2015}.

As a countermeasure against inappropriate content, influence minimization and limitation have been studied. 
However, a challenge with influence minimization methods is that they assume moderation. 
A common challenge for both approaches is the need to consider larger scale networks that include general users. 
To overcome these challenges, this study proposes a method that does not assume moderation and targets a relatively smaller scale network of competition among information disseminators. 
This study focuses on the attention directed toward information disseminators, rather than on information diffusion.
Based on a nonlinear model from previous studies that represents variations in attention to topics~\cite{attention_model}, this study analyzes an attention dynamics model about the attention to information disseminators.
Using the analytical results, this study proposes an ``imitation strategy'' which reduces attention to inappropriate disseminators, derives the conditions under which it works effectively, and verifies its effectiveness through numerical experiments.

The rest of this paper is structured as follows. 
Section 2 introduces the attention dynamics model and describes the analysis using the dimension reduction method. 
Section 3 proposes an imitation strategy based on the analysis results and derives the conditions under which attention can be reduced. 
Section 4 demonstrates the validity of the proposed method through numerical experiments. 
Section 5 concludes the paper. 

\section{Analysis of Attention Dynamics Model}
In this section, I describe the attention dynamics model and discuss the analysis of the reduced system obtained by applying a dimension reduction method.
The objective of this study is to propose a method for mitigating attention to inappropriate content. 
To achieve this, it is first necessary to model the dynamics of attention variation among information disseminators and identify factors that contribute to the reduction of attention.

\subsection{Attention Dynamics Model}
The attention dynamics model incorporates the effect of boredom into the Lotka-Volterra competition equation\cite{attention_model}.
Let the attention $a_i(t)$ of information disseminator $i$ at time $t$, where there are $N$ disseminators $i(i=1,\dots,N)$, be expressed as 
\begin{align}
\displaystyle
    \frac{\dd\, a_i(t)}{\dd\, t}
    =r_i\,a_i(t)\left(1-\frac{b_i(t)+\sum_{j=1}^{N}w_{ij}\,a_j(t)}{K_i}\right).
    \label{competitive}
\end{align}
Here, $r_i>0$ is the natural growth rate, $w_{ij}$ is the competition coefficient with another information disseminator $j$. 
In population fluctuation models of biological species, the degree of overlap in habitats and food sources called niches among different species is considered to determine the intensity of competition~\cite{niche}.  
On social media, the degree of competition can be thought to be determined by how much overlap exists in the target audience, with stronger competition occurring when the content being shared is similar.
$K_i$ represents the carrying capacity, which is the maximum population of biological species that an ecosystem can be sustained. 
In the context of social media, it represents the upper limit of attention that information disseminators can get for a specific genre or topic.
$b_i(t)$ represents the boredom of the viewer and is defined as the convolution of past attention, reflecting that as a disseminator gains more attention, the viewer becomes increasingly bored, 
\begin{align}
    b_i(t)=\int_{0}^{t}a_i(s)\,\exp(-\zeta_i\,(t-s))\dd s.
    \label{boredom}
\end{align}
Here, $\zeta_i>0$ is the boring decay rate, and the closer it is to $0$, the easier it is to get boring. 
Using the Leibniz rule~\cite{Leibniz_rule}, the time derivative of $b_i(t)$ can be written as 
\begin{align}
    \frac{\dd\, b_i(t)}{\dd\, t}
    =a_i(t)-\zeta_i\, b_i(t).
    \label{boredom_td}
\end{align}
In other words, I consider a $2N$-dimensional system of time evolution equations for the attention $\bm{a}=(a_1,\dots,a_N)^{\top}$ and boredom $\bm{b}=(b_1,\dots,b_N)^{\top}$ for $N$ information disseminators. 

\subsection{Analysis based on Spectral Dimension Reduction}
In this section, according to \cite{spectral_dim_red}, I describe overview of the spectral dimension reduction method for network-coupled dynamical systems.

The general form of network-coupled dynamical system is written as 
\begin{align}
    \frac{\dd\, x_i}{\dd\,t} 
    = 
    F(x_i)
    + 
    \sum_{j=1}^N w_{ij}\, G\left(x_i, x_j\right) \quad (i=1,\dots, N),
    \label{nw_dyn}
\end{align}
where $x_i$ is the state variable of node $i$, $F$ is the function related to node $i$ itself, $w_{ij}$ is the edge weight between nodes $i$ and $j$, and $G$ is the interaction function between nodes $i$ and $j$,.
Using a weighted vector $\bm{v}$ that satisfies $\bm{1}^{\top}\bm{v}=1$, let the inner product of $\bm{v}$ and $\bm{x}=(x_1,\dots,x_N)^{\top}$, 
\begin{align}
    R = \bm{v}^{\top}\,\bm{x} = \sum_{i=1}^N v_i\, x_i
\end{align}
represent a one-dimensional observable.
Here, $\bm{1}$ represents an $N$-dimensional vector with all elements equal to $1$.
Choosing $\bm{v}$ as the eigenvector corresponding to the largest eigenvalue of the transpose of the adjacency matrix $\bm{W}^{\top}$ is the optimal approach. 
The time evolution equation of $R$ is written as 
\begin{align}
    \frac{\dd R}{\dd\,t} 
    \approx  
    F(R) + \lambda G(\mu R, R), 
\end{align}
where $\lambda$ is the largest eigenvalue of $\bm{W}^{\top}$ and 
\[
\mu = \frac{\bm{v}^{\top}\,\bm{D}\,\bm{v}}{\lambda\, \bm{v}^{\top}\,\bm{v}}
\] 
where $\bm{D}$ is degree matrix, with node degrees on the diagonal entries. 

The spectral dimension reduction method is applied to the attention dynamics model to obtain a reduced system. 
A simplified model is considered here, where the parameters are set as $r_1=\dots=r_N=r$, $K_1=\dots=K_N=K$, $\zeta_1=\dots=\zeta_N=\zeta$. 
This simplified model implies that information disseminators have equal influence.
First, the simplified attention dynamics model can be arranged in the form of the network-coupled dynamical system \eqref{nw_dyn}, as shown in
\begin{align}
    \frac{\dd\, a_i(t)}{\dd\, t}
    =&
    r\,a_i(t)
    \left(
    1-\frac{\int_{0}^{t}a_i(s)\,\exp(-\zeta\,(t-s))\dd s}{K}
    \right)\notag \\
    &+ 
    \sum_{j=1}^N w_{ij}\,\frac{r\,a_i(t)\,a_j(t)}{K}. 
\end{align}
The first term of right-hand side is the function $F$ related to node $i$ and the second term includes $G$ which pertains to the interaction between node $i$ and $j$. 
Let the one-dimensional observable be 
\begin{align}
    A = \bm{v}^{\top}\, \bm{a}. 
    \label{A}
\end{align}
The time evolution of $A$ can be written as 
\begin{align}
    \frac{\dd\, A}{\dd\,t} 
    &\approx 
    F(A) + \lambda\, G(\mu A, A)\notag\\
    &= 
    rA\left(1-\frac{\int_{0}^{t}A(s)\,\exp(-\zeta\,(t-s))\dd s}{K}\right) 
    + 
    \lambda \frac{r\mu A^2}{K}. 
    \label{reduced_A}
\end{align}
Here, let 
\[
B=\int_{0}^{t}A(s)\,\exp(-\zeta\,(t-s))\dd s.
\] 
Similarly to the case of node-specific boredom effect \eqref{boredom_td}, the time evolution of $B$ can be expressed as 
\begin{align}
    \frac{\dd\,B}{\dd\,t} 
    = A - \zeta B.
    \label{reduced_B}
\end{align}
Based on the above, a two-dimensional reduced system \eqref{reduced_A} and \eqref{reduced_B} can be derived from the $2N$-dimensional attention dynamics model \eqref{competitive} and \eqref{boredom_td}. 

From the reduced system \eqref{reduced_A} and \eqref{reduced_B}, the relationship 
\begin{align}
A = \frac{K\zeta}{1+\zeta \mu \lambda} 
\label{reduced_relation_attention}
\end{align}
and
\begin{align}
B = \frac{K}{1+\zeta \mu \lambda}
\label{reduced_relation_boredom}
\end{align}
holds at the fixed point. 
The equation \eqref{reduced_relation_attention} indicates that the degree of attention $A$ decreases as the largest eigenvalue $\lambda$ increases.
Therefore, I have identified that increasing the largest eigenvalue is a key factor in mitigating attention.
Since the largest eigenvalue is determined by the network structure, specifically the competitive relationships among information disseminators, modifying the competitive dynamics can effectively mitigate attention.

\section{Imitation Strategy for Mitigating Attention}\label{proposal}
\subsection{Imitation Strategy}
In this section, I propose a method to reduce attention to inappropriate content, based on the relationship \eqref{reduced_relation_attention} derived from the analysis in the previous section. 
Based on this relationship, a greater largest eigenvalue of the transposed adjacency matrix results in less attention.
From this fact, I develop a strategy to increase the largest eigenvalue through the actions of a specific information disseminator. 
The largest eigenvalues tend to increase as the average edge weight and edge density become larger. 
The adjacency matrix represents a competition network among information disseminators, where the edge weights correspond to competitive coefficients, indicating the strength of competition; in other words, how much they disseminate information on similar topics.
Taking these points into account, an information disseminator can increase edge weights by imitating another disseminator, thereby raising the largest eigenvalue and ultimately reducing the degree of attention. 
Let this strategy be called the imitation strategy, which I describe below. 
As the edge weight is determined by the similarity of the information disseminators' content, I consider a feature vector. 
For example, methods of embedding that represents a sentence as a vector can be used for this purpose~\cite{doc2vec,user2vec}. 
Let the feature vectors of the $N$ information disseminators be $\bm{c}_1, \bm{c}_2, \dots, \bm{c}_N$. 
Similarity of information disseminators $i$ and $j$ is calculated as cosine similarity, 
\begin{align}
    S_{ij} = \frac{\bm{c}_i^{\top}\,\bm{c}_j}{\sqrt{\bm{c}_i^{\top}\,\bm{c}_i}\,\sqrt{\bm{c}_j^{\top}\,\bm{c}_j}}. 
\end{align}
The cosine similarity $S$ has values in $[-1, 1]$. 
When $S<0$, they disseminate unrelated content each other and adjacency matrix is determined as 
\begin{align}
    \bm{W}_{ij} = 
    \begin{cases}
        \max\left(S_{ij}, 0\right) & i\neq j\\
        0 & i=j
    \end{cases}. 
\end{align}
The largest eigenvalues tend to increase when the edge density is high, and they also tend to increase when the average edge weight is large. 
Using this property, I propose an imitation strategy in which an imitation node imitates the content disseminated by a target node to intensify competition, increase edge weights, and thereby increase the largest eigenvalue, aiming to reduce the overall attention. 
Hereafter, the subscript for imitator nodes is denoted by $\iota$, and the subscript for target nodes is denoted by $\tau$.
Through imitation, the feature vector of the imitation node evolves to match that of the target node as follows: 
\begin{align}
    \hat{\bm{c}}_{\iota} = \bm{c}_{\tau}.
\end{align}
The changes in feature vectors caused by imitation alter the competitive relationships between nodes, resulting in the adjacency matrix becoming 
\begin{align}
    \hat{\bm{W}}_{ij} = 
    \begin{cases}
        \max ( S(\hat{\bm{c}}_{\iota}, \bm{c}_j), 0) & i=\iota \text{ and } j\neq \iota\\
        \max ( S(\bm{c}_i, \hat{\bm{c}}_{\iota}), 0) & i\neq \iota \text{ and } j=\iota\\
        \bm{W}_{ij} & \text{otherwise}.
    \end{cases}
\end{align}
Let the largest eigenvalue of the adjacency matrix after the change be denoted as $\hat{\lambda}$. 
The above outlines the procedure of the imitation strategy.
This strategy increases the edge weight between the imitator and the target; however, since it also affects the edge weights between the imitator and other nodes, success is not always guaranteed.
Therefore, in the next subsection, I derive the success condition of the imitation strategy and determine the optimal selection of the imitator and target.

\subsection{Success condition of imitation strategy}
The imitation strategy is considered successful if the change in the largest eigenvalue satisfies $\hat{\lambda}-\lambda>0$. 
Here, the change in the largest eigenvalue resulting from the imitation strategy is evaluated using matrix perturbation theory \cite{matrix_perturbation}.
Let the matrix representing the change in the adjacency matrix be denoted as $\bm{P}$. 
Then, the adjacency matrix after the change can be decomposed as 
\begin{align}
    \hat{\bm{W}} = \bm{W} + \bm{P}.
\end{align}
Introducing the perturbation coefficient $\epsilon$, the adjacency matrix with perturbation can be expressed as 
\begin{align}
    \hat{\bm{W}}(\epsilon) = \bm{W} + \epsilon\, \bm{P}. 
\end{align}
The perturbation matrix $\bm{P}$ has non-zero values only in the off-diagonal elements of the $\iota$-th row and the $\iota$-th column, and can be expressed as 
\begin{align}
    P_{ij} = 
    \begin{cases}
        p_{\iota} & \text{if}\, (i=\iota \,\text{and}\, j\neq \iota) \,\text{or}\, (i\neq \iota \,\text{and}\, j = \iota)\\
        0 & \text{otherwise}. 
    \end{cases}
\end{align}
To prepare for later use, the eigenvalues and eigenvectors of the matrix $\bm{P}$ are calculated.
The eigenvalues and eigenvectors of the matrix $\bm{P}$ can be analytically derived by performing the cofactor expansion of the characteristic polynomial along the $\iota$-th row. 
Let $M_{\iota j}$ denote the $(\iota, j)$ minor of $\psi\, \bm{I} - \bm{P}$ and let $M_{kl}^{(\iota j)}$ denote the $(k,l)$ minor of $M_{\iota j}$. 
The characteristic polynomial can be calculated as 
\begin{align}
    &\det(\psi\, \bm{I} - \bm{P}) \notag\\
    &=\sum_{j=1, j\neq i}^{n} p_j\, (-1)^{\iota+j}\,M_{\iota j} + \psi (-1)^{2\iota}M_{\iota \iota}\notag\\
    &=\sum_{j=1}^{i-1} p_j\, (-1)^{\iota+j}\,p_j\,(-1)^{\iota+j-1}M_{j\,\iota-1}^{(\iota j)}\notag\\ 
    &\quad + \sum_{j=\iota+1}^{n} p_j\, (-1)^{\iota+j}\,p_j\,(-1)^{j-1+\iota}M_{j-1\,\iota}^{(\iota j)} 
    +\psi (-1)^{2\iota}\psi^{n-1}\notag\\
    &=\sum_{j=1, j\neq i}^{n} p_j^2\,(-1)^{2(\iota+j)-1}\psi^{n-2} + \psi^{n}(-1)^{2\iota}\notag\\
    &=\psi^{n-2}\left(\psi^2 - \sum_{j=1, j\neq \iota}^{n}p_j^2\right). 
\end{align}
Here, $M_{\iota \iota}$ turns out to be an $(n-1)$-dimensional square matrix given by $M_{\iota \iota} = \mathrm{diag}(\psi, \dots, \psi)$.
Furthermore, when $j < \iota$, the $(j, \iota-1)$ minor $M_{j\,\iota-1}^{(\iota j)}$ of $M_{\iota j}$ is the determinant of a matrix with $\psi$ appearing on the diagonal elements, and when $j > \iota$, the $(j-1, \iota)$ minor $M_{j-1\,\iota}^{(\iota j)}$ of $M_{\iota j}$ is the determinant of a matrix with $\psi$ appearing on the diagonal elements.
Thus, the eigenvalues are $\pm \sqrt{\sum_{j=1, j\neq \iota}^{n}p_j^2}$ and $0$ with a multiplicity of $n-2$. 
The non-zero eigenvalues are denoted as 
\begin{align}
    \psi_1 &= -\sqrt{\sum_{j=1, j\neq \iota}^{n}p_j^2}
\end{align}
and 
\begin{align}
    \psi_n &= \sqrt{\sum_{j=1, j\neq \iota}^{n}p_j^2}. 
\end{align}
The eigenvectors corresponding to the eigenvalues $\psi_1$ and $\psi_n$ can each be expressed as 
\begin{align}
    \phi_1(j) 
    = 
    \begin{cases}
    \displaystyle
        \frac{p_j}{\psi_1} & \text{if}\, j\neq \iota\\
        \displaystyle
        1 & \text{if}\, j=\iota
    \end{cases}
    \label{eigenvector_phi_1}
\end{align}
and 
\begin{align}
    \phi_n(j) 
    = 
    \begin{cases}
    \displaystyle
        \frac{p_j}{\psi_n} & \text{if}\, j\neq \iota\\
        \displaystyle
        1 & \text{if}\, j=\iota.
    \end{cases}
    \label{eigenvector_phi_n}
\end{align}
These eigenvectors have their $\iota$-th components related by a sign inversion, while the other components are identical.

The largest eigenvalue of the adjacency matrix of a network altered by imitation strategies can be expressed as 
\begin{align}
    \hat{\lambda} \approx \lambda + \epsilon\, \bm{v}^{\top} \bm{P} \bm{v}
    \label{perturbation}
\end{align}
using matrix perturbation theory~\cite{matrix_perturbation}.
Imitation strategies are effective if the eigenvalue increases; therefore, this strategy works well if the second term on the right-hand side of equation \eqref{perturbation} takes a positive value.
Let the normalized eigenvector of $\bm{\phi}$ be denoted as $\bm{q}$, and the matrix formed by arranging these vectors be denoted as $\bm{Q}$.
Additionally, let $\bm{\Psi}$ be the matrix with the eigenvalues of $\bm{P}$ arranged along its diagonal.
Then, the second term on the right-hand side of equation \eqref{perturbation} can be rewritten as 
\begin{align}
    \bm{v}^{\top} \bm{P} \bm{v} 
    &=(\bm{Q}^{\top}\bm{v})^{\top}\bm{\Psi}(\bm{Q}^{\top}\bm{v})\notag\\
    &=\psi_n ((\bm{q}_n^{\top}\bm{v})^{2}-(\bm{q}_1^{\top}\bm{v})^{2}).
    \label{perturbation_term_simplify}
\end{align}
Here, I used the fact that $\psi_1 = -\psi_n$ and that the other eigenvalues of the perturbation matrix $\bm{P}$ are $0$.
From the equation \eqref{perturbation_term_simplify}, It can be seen that the largest eigenvalues increase if condition 
\begin{align}
    (\bm{q}_n^{\top}\bm{v}+\bm{q}_1^{\top}\bm{v})(\bm{q}_n^{\top}\bm{v}-\bm{q}_1^{\top}\bm{v})&>0
\end{align}
is satisfied.
By using the fact that the eigenvectors $\bm{q}_1$ and $\bm{q}_n$ have the same values for all components except the $\iota$-th component, where they have a sign-reversed relationship, the condition can be written as 
\begin{align}
    (q_n(\iota)\,v(\iota))
    \left(\sum_{j=1, j\neq \iota}^{n}q_n(j)\,v(j)\right)
    >0
    \label{condition_fully_simplified}
\end{align}
The following discussion examines the derived condition \eqref{condition_fully_simplified}.
First, in the condition \eqref{condition_fully_simplified}, $ q_n(\iota)\,v(\iota) > 0 $ always holds.
Therefore, the largest eigenvalues increase if 
\begin{align}
\sum_{j=1, j\neq \iota}^{n}q_n(j)\,v(j) > 0
\label{success_condition}
\end{align}
is satisfied.
Since all components of the eigenvector $\bm{v}$ corresponding to the largest eigenvalue of the adjacency matrix $\bm{W}$ are non-negative, the increase or decrease of the eigenvalue is determined by the normalized change in edge weights $q_j$ $(j=1, \dots, n, j \neq \iota)$.
Increasing the edges weight between the imitator and other nodes leads to a higher largest eigenvalue, so imitation that maximizes competition is desirable.
The components of the eigenvector $\bm{v}$ corresponding to the largest eigenvalue of the adjacency matrix represent the importance of the nodes in the network and are known as the eigenvector centrality~\cite{eig_cent_1,eig_cent_2}.
Eigenvector centrality is a metric that differs from degree centrality, which directly considers the number of edges as a measure of a node's importance. 
Instead, eigenvector centrality assigns higher importance to nodes that are connected to other highly important nodes.
In the competitive network considered in this study, eigenvector centrality implies that nodes competing with highly competitive nodes are also considered more important.
Imitating highly competitive nodes increases competition not only with the target node but also with its neighboring nodes. 
Therefore, equation \eqref{success_condition} is a reasonable condition for the success of the imitation strategy.
Additionally, when the left-hand side takes a positive value, the presence of $v(\iota)$ on the left-hand side indicates that the intervention effect is greater for node $\iota$ if it has higher eigenvector centrality.
Note that the impact is significant not only when the largest eigenvalue increases but also when it decreases.

\section{Numerical Experiments}
\subsection{impact of eigenvector centrality on imitation strategy}
In this section, I verify the effectiveness of the insights obtained in Section \ref{proposal} through numerical experiments and clarify the conditions under which the imitation strategy works effectively.
To achieve this, I conduct numerical experiments using three networks with different characteristics.
The networks consist of a relatively sparse network with low edge density, a dense network with high edge density, and a heterogeneous network with partially dense regions.
Conducting experiments on networks with relatively low and high edge densities aims to evaluate the impact of competition increase or decrease on the success of the imitation strategy.
In other words, this approach facilitates the assessment of the effects of introducing new competition and reducing existing competition.
Additionally, experiments on heterogeneous networks are conducted to enable a more effective evaluation of the impact of eigenvector centrality, as certain nodes exhibit high eigenvector centrality in heterogeneous networks.
For each network, I apply the imitation strategy to all combinations of imitator and target nodes, and evaluate the change in the largest eigenvalue.
I also examine the influence of eigenvector centrality on the results.

\begin{figure}[tb]
    \centering
    \includegraphics[width=0.5\linewidth]{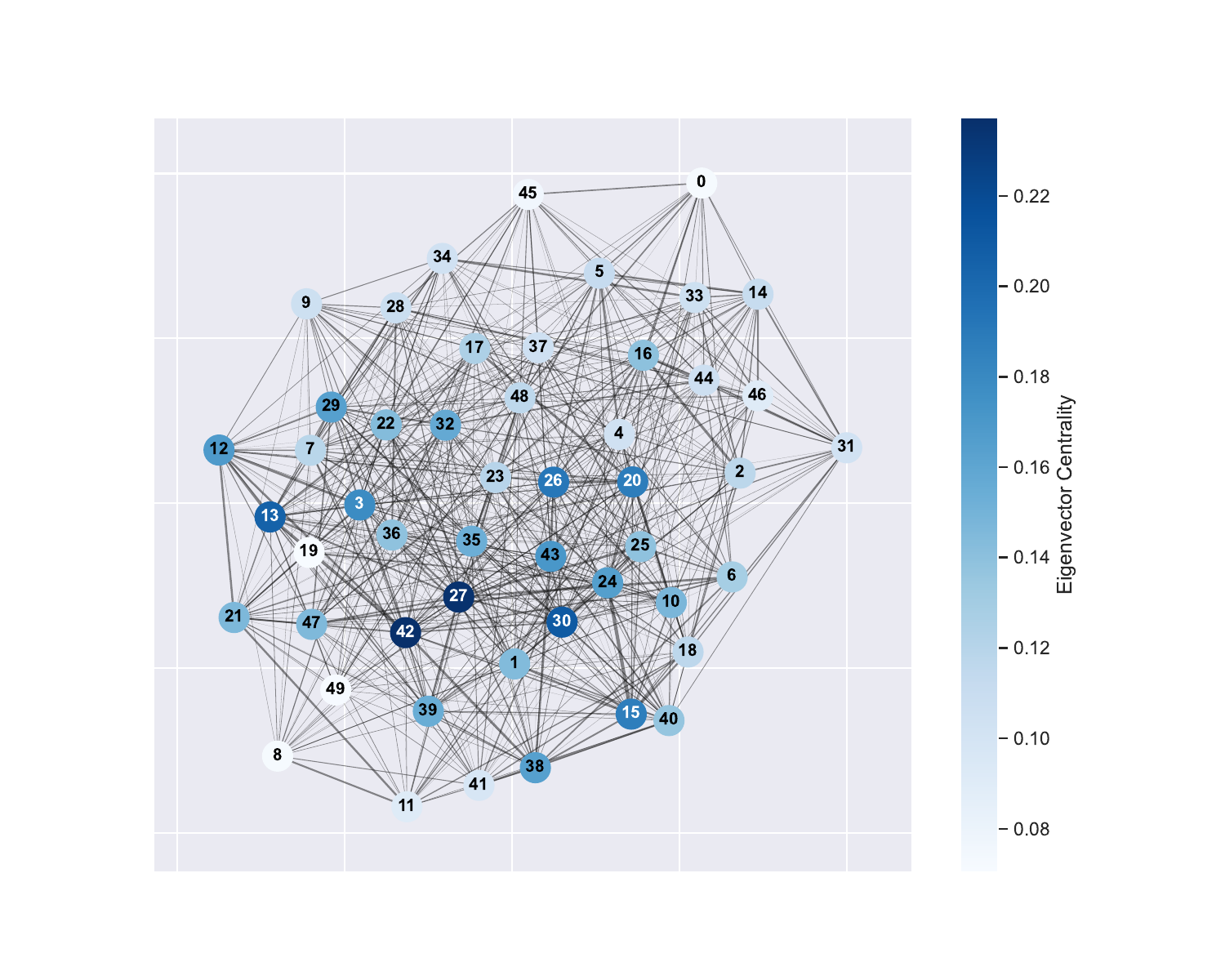}
    \caption{A sparse network and the eigenvector centrality of each node}
    \label{fig:sparse_eigen_nw}
\end{figure}
\begin{figure}[tb]
    \centering
    \includegraphics[width=0.5\linewidth]{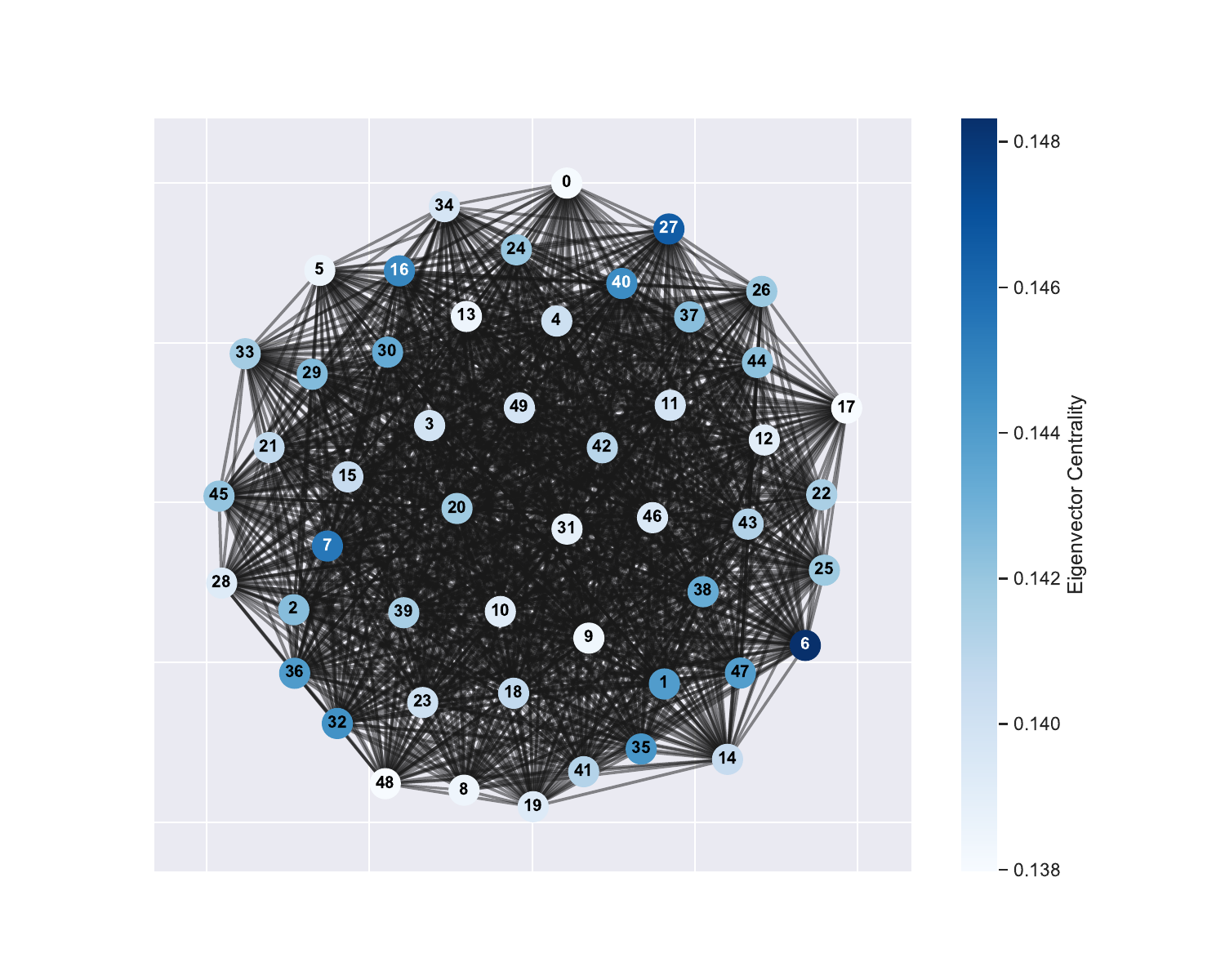}
    \caption{A dense network and the eigenvector centrality of each node}
    \label{fig:dense_eigen_nw}
\end{figure}
\begin{figure}[tb]
    \centering
    \includegraphics[width=0.5\linewidth]{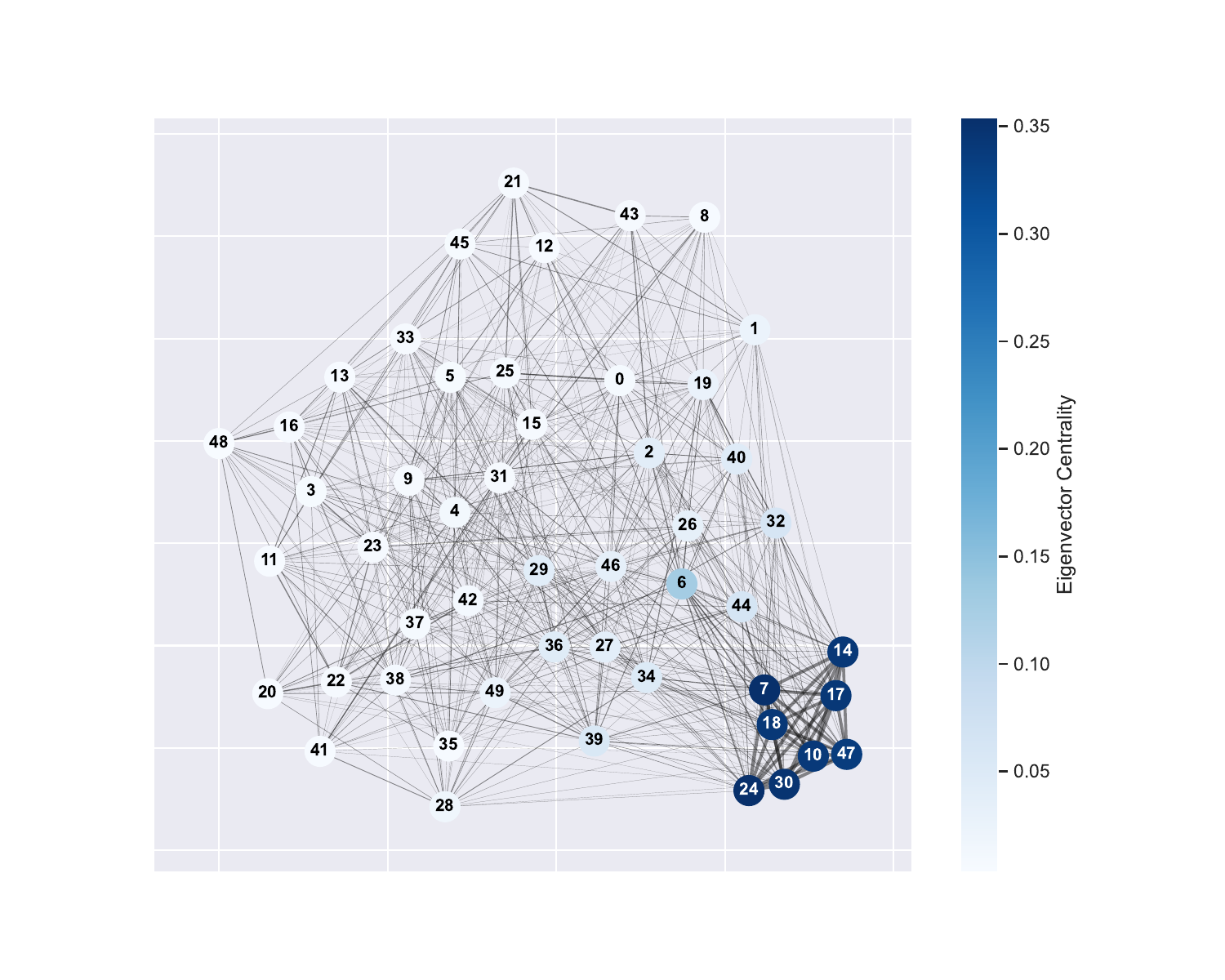}
    \caption{A heterogeneous network and the eigenvector centrality of each node}
    \label{fig:heterogeneous_eigen_nw}
\end{figure}
First, I describe the method for generating networks with distinct edge densities based on node feature vectors.  
This is achieved by probabilistically generating the feature vectors.  
In the numerical experiments, the feature vectors are assumed to have $100$ dimensions and are normalized to have a norm of $1$.  
To generate a sparse network, each element of the feature vectors is generated as a uniform random number in the range of $[-1, 1]$.  
For a dense network, each element of the feature vectors is generated as a uniform random number in the range of $[0, 1]$.  
To create a heterogeneous network, the feature vectors of some nodes are generated as uniform random numbers in the range of $[0, 1]$, while the remaining nodes' feature vectors are generated in the range of $[-1, 1]$.  
After generating the feature vectors, they are normalized to have a norm of $1$.  
Once the feature vectors for all nodes are generated, the network can be constructed by calculating the cosine similarity between vectors and determining edge weights accordingly.  
The networks generated by this method are visualized in the figures.  
Figure \ref{fig:sparse_eigen_nw} shows the sparse network, Figure \ref{fig:dense_eigen_nw} shows the dense network, and Figure \ref{fig:heterogeneous_eigen_nw} represents the heterogeneous network.  
Additionally, the color of each node indicates its eigenvector centrality, with darker blue representing nodes with higher centrality.  

\begin{figure}[tb]
    \centering
    \includegraphics[width=0.5\linewidth]{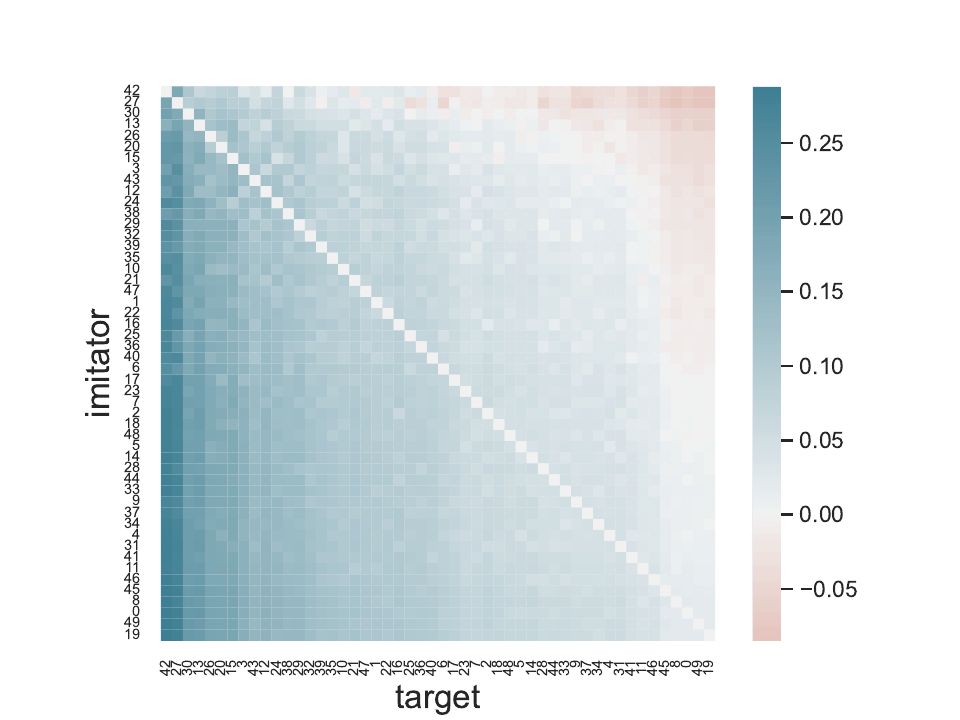}
    \caption{Change in the largest eigenvalue for each combination of imitator and target in a sparse network}
    \label{fig:sparse_eigen}
\end{figure}

The experimental results for the sparse network are shown in Figure \ref{fig:sparse_eigen}.
The vertical axis represents the imitator nodes, while the horizontal axis represents the target nodes, both sorted in descending order of eigenvector centrality.
The change in eigenvalue is visualized as a heatmap, where darker blue indicates a larger increase, and darker red indicates a larger decrease.
The results show that selecting nodes with high eigenvector centrality as target nodes is more effective.
When nodes with high eigenvector centrality are chosen as imitators and nodes with low centrality are selected as targets, the largest eigenvalue decreases.
This is due to abandoning existing competition.
In the case of sparse networks, since the original level of competition is relatively mild, the region where the largest eigenvalue decreases is small.

\begin{figure}[tb]
    \centering
    \includegraphics[width=0.5\linewidth]{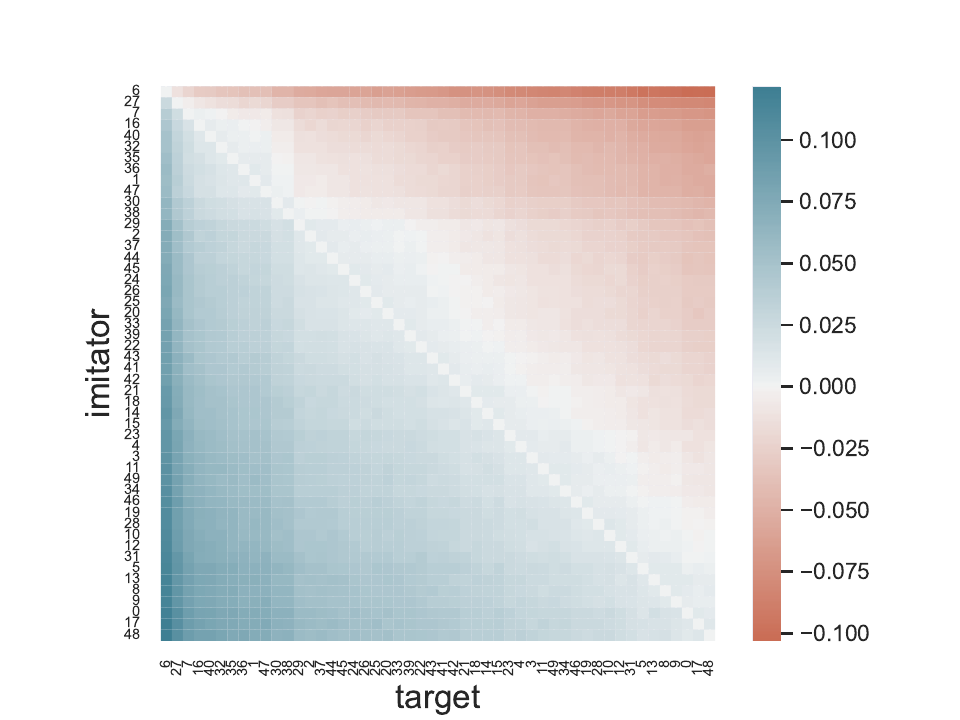}
    \caption{Change in the largest eigenvalue for each combination of imitator and target in a dense network}
    \label{fig:dense_eigen}
\end{figure}

The experimental results for the dense network are shown in Figure \ref{fig:dense_eigen}.
As in the experiment with the sparse network, the nodes are sorted in descending order of eigenvector centrality.
Selecting nodes with high eigenvector centrality as target nodes is effective, similar to the case of the sparse network.
However, it can be observed that the region where the largest eigenvalue decreases has expanded.
The reason for this is that in dense networks, the original level of competition is intense, and in many cases, the competition that is abandoned has a greater impact than the newly generated competition through imitation.
It is generally observed that the imitation strategy fails when the imitator has higher eigenvector centrality than the target.

\begin{figure}[tb]
    \centering
    \includegraphics[width=0.5\linewidth]{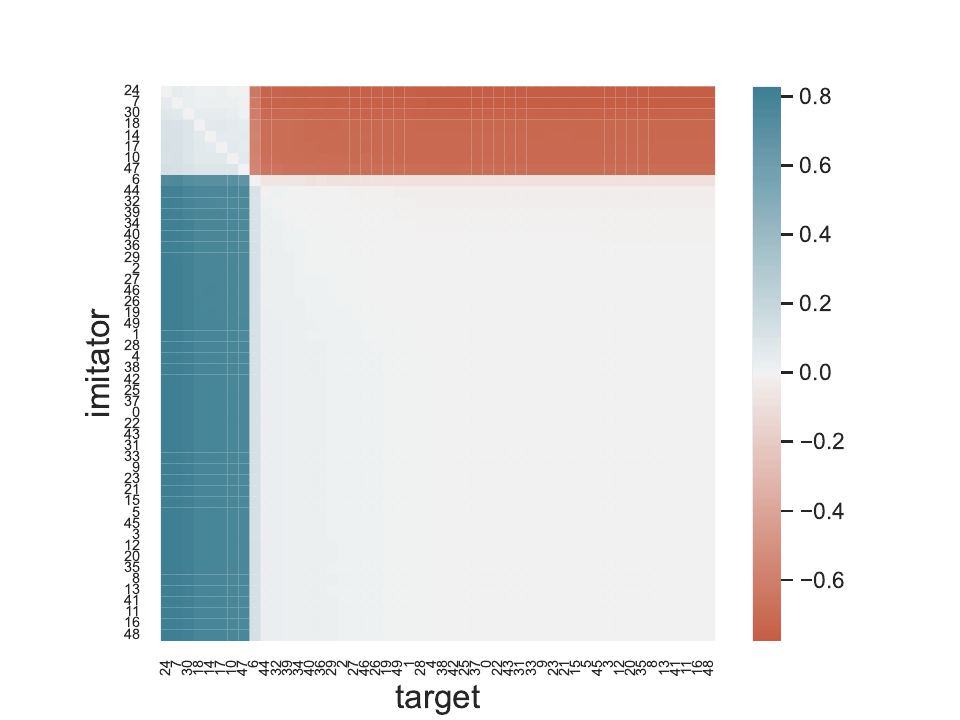}
    \caption{Change in the largest eigenvalue for each combination of imitator and target in a heterogeneous network}
    \label{fig:heterogeneous_eigen}
\end{figure}

Finally, the experimental results for the heterogeneous network, where some nodes compete intensely, are shown in Figure \ref{fig:heterogeneous_eigen}.
As in the other two experiments, the nodes are sorted in descending order of eigenvector centrality.
Selecting a node with high eigenvector centrality as a target node has a significant effect on increasing the largest eigenvalue.
On the other hand, selecting some of the nodes with high eigenvector centrality as an imitator and choosing a target from nodes with lower centrality results in a decrease in the largest eigenvalue, indicating that the strategy does not work well.
Additionally, selecting an imitator and a target from nodes with low centrality does not significantly affect the largest eigenvalue.

From the above experimental results, it was found that both minimizing the reduction of existing competition and generating new competition are important for the success of the imitation strategy.  
Furthermore, it was revealed that eigenvector centrality of the competition networks plays a key role.

\subsection{Evaluation of Imitation Strategy under Noisy Conditions}
In this section, I evaluate the effectiveness of the imitation strategy under noisy conditions. 
Since it is common for the target node's information to be incomplete or for the imitation process to be imperfect, I examine how effective the proposed method remains when only imperfect imitation is possible.

The evaluation experiment is conducted as follows. 
I generated ten different patterns for each type of network—sparse, dense, and heterogeneous density—using the same method as in the previous section. 
The imitation strategy was applied to all possible pairs of imitators and targets. 
However, instead of performing perfect imitation, I added noise following a normal distribution to each component of the feature vector. 
The mean of the normal distribution was set to $0$, and the standard deviation varied from $0$ to $1$ in increments of $0.1$, resulting in ten different conditions, including the noiseless case.

\begin{figure}
    \centering
    \includegraphics[width=0.5\linewidth]{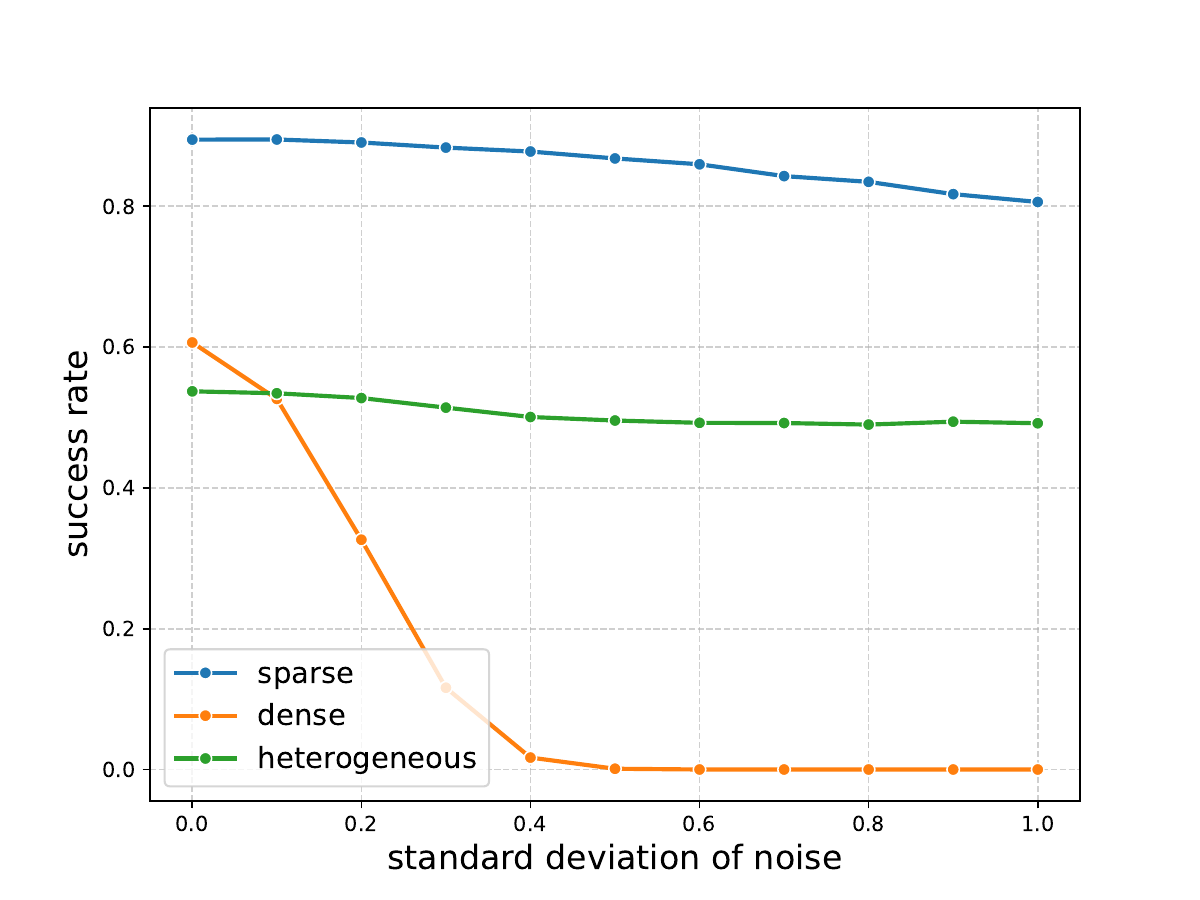}
    \caption{Success rate of the imitation strategy under varying noise level}
    \label{fig:noise}
\end{figure}

For each experiment, I observed the changes in largest eigenvalues for all imitator-target pairs. 
A case was considered successful if the eigenvalue increased, and I evaluated the proportion of successful cases. 
A higher number of successful cases suggests greater applicability. 
The experimental results are shown in Figure \ref{fig:noise}.
In the case of sparse networks, the success rate gradually decreased as noise increased. 
This is because, in sparse networks where competition is relatively mild, the decrease in existing competition due to imitation has little impact. 
As a result, even with noisy imitation, it still sufficiently contributes to increasing competition.
For dense networks, the success rate declined with increasing noise, and failure was inevitable when the standard deviation exceeded $0.5$. 
The sharp decrease in success rate in dense networks can be attributed to the fact that the reduction in existing strong competition due to imitation outweighs the increase in new competition introduced by the imitation strategy.
In heterogeneous networks, the success rate remained almost unchanged. 
This is because successful cases are likely to occur when competition in the dense subnetwork is maintained while competition in sparse the subnetwork increases. 
Therefore, even with noisy imitation, the success rate was not significantly affected.

\section{Conclusion}
In this paper, I proposed an imitation strategy to suppress attention to inappropriate content that spreads under the attention economy.
Additionally, I derived the conditions under which the imitation strategy works effectively and evaluated them through numerical experiments using three types of networks with different edge density characteristics.
As a result, it was found that the imitation strategy works effectively when imitators and targets are selected based on eigenvector centrality.

Some limitations exist in this study.
First, it deals with a simplified model that assumes all information disseminators have equal influence.
It is necessary to conduct an analysis using a more general model that considers variations in the influence of information disseminators.
Second, the proposed method does not consider whether nodes are disseminating inappropriate content.
To develop a method that integrates proposed method with existing inappropriate content detection techniques, it is essential to consider that the impact of imitation extends beyond inappropriate disseminators, making it necessary to maximize the reduction of attention to inappropriate disseminators while minimizing its impact on regular disseminators who share legitimate content.
Third, validation using real-world data is necessary.
To conduct such validation, it is essential to estimate feature vectors from the content posted by information disseminators and understand the structure of the competitive network.
This involves multiple challenges, including identifying key disseminators within a specific topic, collecting relevant data, developing methods for estimating feature vectors, and integrating the proposed approach with existing techniques, making it an important issue to be addressed in future work.

When implementing the proposed method in real-world applications, it is necessary to understand the competitive relationships among information disseminators and either encourage disseminators to post imitation content or have the platform preemptively prepare influential accounts.
Additionally, it is necessary to adapt to changes in the network structure, such as the entry and exit of information disseminators.
Furthermore, in the near future, inappropriate content may be predominantly generated automatically by generative artificial intelligence. To counter this, research on the automatic creation of imitation content may also become necessary.

\bibliographystyle{acm}

\end{document}